\documentclass[a4paper,11pt]{article}
\usepackage{pos}
\usepackage{lineno}
\title{The AGILE real-time analysis pipelines in the multi-messenger era}
\ShortTitle{AGILE RTA pipeline}

\author*[a,b]{N. Parmiggiani}
\author[a]{A. Bulgarelli}
\author[c]{A. Ursi}
\author[a]{V. Fioretti}
\author[a]{L. Baroncelli}
\author[a]{A. Addis}
\author[a]{A. Di Piano}
\author[d,e]{C. Pittori}
\author[d,e]{F. Verrecchia}
\author[d,e]{F. Lucarelli}
\author[d]{M. Tavani}
\author[b]{D. Beneventano}

\affiliation[a]{INAF/OAS Bologna, Via P. Gobetti 93/3, I-40129 Bologna, Italy.}

\affiliation[b]{Universit\`{a} degli Studi di Modena e Reggio Emilia, DIEF - Via Pietro Vivarelli 10, I-41125 Modena, Italy.}

\affiliation[c]{INAF/IAPS Roma, Via del Fosso del Cavaliere 100, I-00133 Roma, Italy.}

\affiliation[d]{INAF/OAR Roma, Via Frascati 33, I-00078 Monte Porzio Catone, Roma, Italy.}

\affiliation[e]{ASI/SSDC Roma, Via del Politecnico snc, I-00133 Roma, Italy.}

\emailAdd{nicolo.parmiggiani@inaf.it}

\abstract{In the multi-messenger era, space and ground-based observatories usually develop real-time analysis (RTA) pipelines to rapidly detect transient events and promptly share information with the scientific community to enable follow-up observations. These pipelines can also react to science alerts shared by other observatories through networks such as the Gamma-Ray Coordinates Network (GCN) and the Astronomer's Telegram (ATels). AGILE is a space mission launched in 2007 to study X-ray and gamma-ray phenomena. This contribution presents the technologies used to develop two types of AGILE pipelines using the RTApipe framework and an overview of the main scientific results. The first type performs automated analyses on new AGILE data to detect transient events and automatically sends AGILE notices to the GCN network. Since May 2019, this pipeline has sent more than 50 automated notices with a few minutes delay since data arrival. The second type of pipeline reacts to multi-messenger external alerts (neutrinos, gravitational waves, GRBs, and other transients) received through the GCN network and performs hundreds of analyses searching for counterparts in all AGILE instruments' data. The AGILE Team uses these pipelines to perform fast follow-up of science alerts reported by other facilities, which resulted in the publishing of several ATels and GCN circulars.}

\FullConference{37$^{\rm{th}}$ International Cosmic Ray Conference (ICRC 2021)\\
		July 12th -- 23rd, 2021\\
		Online -- Berlin, Germany}

\begin{document}
\maketitle

\section{Introduction} \label{sec:intro}

AGILE (Astrorivelatore Gamma ad Immagini LEggero - Light Imager for Gamma-Ray Astrophysics) is a scientific mission of the Italian Space Agency (ASI) launched on 23rd Apr 2007. The AGILE payload consists of the Silicon Tracker (ST), the SuperAGILE X-ray detector, the CsI(Tl) Mini-Calorimeter (MCAL), and an AntiCoincidence System (ACS). The combination of ST, MCAL, and ACS form the Gamma-Ray Imaging Detector (GRID) \cite{2009A&A...502..995T}.

\begin{figure*}[!htb]
	\centering
	  \includegraphics[width=\linewidth]{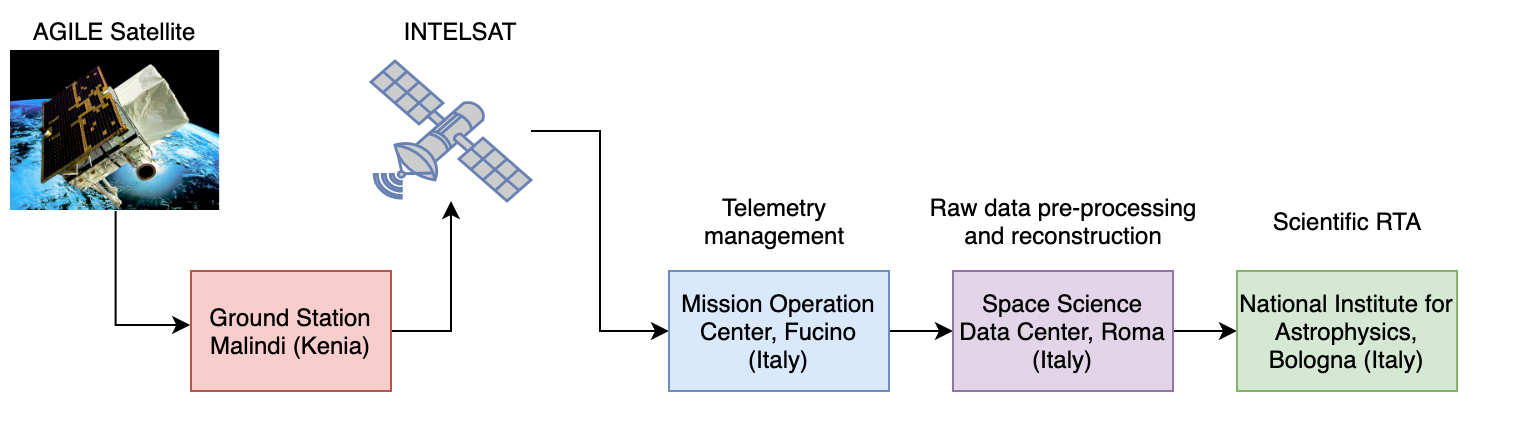}
	\caption{Schema of the AGILE data flow, from the satellite to the scientific pipelines.}
	\label{fig:data_flow}
\end{figure*}

Fig. \ref{fig:data_flow} shows how the raw telemetry data produced by the AGILE instrument are downlinked in the ASI ground station (Malindi, Kenya), and this may happen at each orbit (about every 90 minutes) or according to the ground station availability schedule. The contact packet contains the data of one or more satellite orbits. The data are immediately transferred to Italy via a dedicated ASINet network to Telespazio, Fucino (Italy), and then to the ASI Space Science Data Center (\url{https://www.ssdc.asi.it}) (SSDC) \cite{2019RLSFN.tmp...23P}. The automatic pipeline system developed at SSDC performs the preprocessing (by the AGILE Preprocessing System), reducing, archiving, and distributing the data. The reconstructed data are then sent to the National Institute for Astrophysics (INAF/OAS) in Bologna (Italy) for further scientific analysis, performed by the automated software described in this contribution. The AGILE data flow and the reconstruction software systems are described in detail in \cite{2014ApJ...781...19B,2019ExA....48..199B}.

In Multi-Wavelength (MW) astronomy, astrophysical phenomena are studied with facilities that collect electromagnetic radiation at different wavelengths (gamma-ray, x-ray, ultraviolet, optical, infrared, radio, etc.). In recent years, the astrophysical landscape changed due to two main transient events. The first event was the Gamma-Ray Burst (GRB) detected on 17-Aug-2017 by the Fermi Gamma-ray Space Telescope \cite{2018ApJ...861...85A} after the LIGO-Virgo Collaboration detection of the Gravitational Wave (GW) GW170817, produced during a binary neutron star coalescence \cite{2017PhRvL.119p1101A}. This GW event was observed with multi-messenger instruments (gravitational and electromagnetic waves detectors) during the following days \cite{2017ApJ...848L..12A}. The second event was a neutrino event detected by IceCube (IceCube-170922A), which led to an extensive campaign of observations \cite{2018Sci...361.1378I}, during which the Fermi-LAT instrument detected an enhanced gamma-ray emission from the blazar TXS 0506+056 with time coincidence. These events started the so-called "Multi-Messenger era" (MM) \cite{2019NatRP...1..585M}. AGILE contributed to this new astronomy era \cite{2019RLSFN.tmp...38T} also thanks to its improved reaction capability to short timescale transients obtained with the real-time analysis (RTA) software developed in the context of MW and MM astronomy and described in this contribution. 

A science alert is a communication from/to the astrophysical community that a transient phenomenon occurs in the sky. In MW and MM astronomy, observatories share science alerts to study the same physical phenomena with different "messenger" signals (electromagnetic radiation, gravitational waves, and neutrinos) through communication networks such as the Gamma-Ray Coordinates Network (\url{https://gcn.gsfc.nasa.gov}) (GCN) and The Astronomer's Telegram (\url{http://www.astronomerstelegram.org}) (ATel). The GCN is a low-latency system (less than 30 sec) and is used by automated software for fast reaction to science alerts. This service sends two types of communications: (i) circulars that are written in a human-readable format and (ii) notices that are designed to be interpreted by software.  

Usually, data analysis pipelines are developed for space and ground-based projects to identify as fast as possible transient phenomena (e.g. GRBs), send science alerts to the astrophysical community, and speed up the reaction time to science alerts sent by others facilities. The RTApipe framework \cite{2021ADASSParmiggiani} has been designed to help the development of the RTA pipelines for high-energy projects and fulfil the requirement of the MM and MW astronomy context. 

The RTA pipelines developed by the AGILE Team can satisfy two main use cases: (i) they react to an external science alert, received through the GCN network, and start scientific analysis on available data or wait for data acquisition in the time interval of interest; (ii) they execute periodical scientific analysis searching for transient events inside the data acquired by AGILE to generate science alerts and sent them to the scientific community. These two use cases allow the AGILE Team to perform a fast follow-up of transient events detected by AGILE or other observatories. The AGILE Team has an astronomer on-duty who is in charge of the follow-up of science alerts. There are duty shifts dedicated to this activity.

The architecture of the AGILE RTA system is illustrated in Section \ref{sec:architecture}. The pipelines developed for AGILE and the main results obtained are described in Section \ref{sec:agilepipe}. Finally, in Section \ref{sec:conclusion} the conclusions are reported.

\section{AGILE real-time analysis software architecture} \label{sec:architecture}

\begin{figure*}[!htb]
	\centering
	  \includegraphics[width=\linewidth]{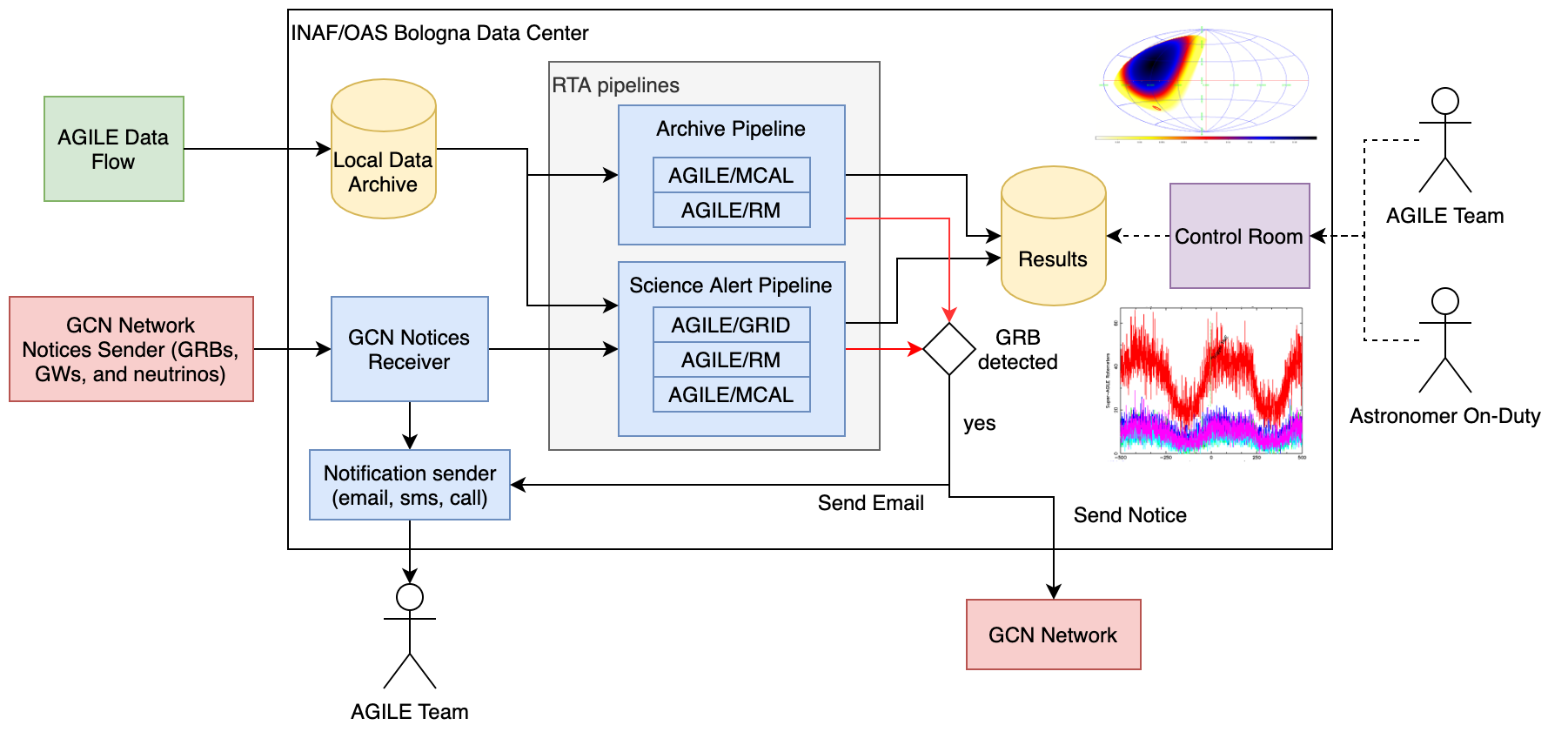}
	\caption{Architecture of the AGILE real-time analysis software system.}
	\label{fig:architecture}
\end{figure*}

Fig. \ref{fig:architecture} shows the general context of the AGILE RTA system and the components of the RTA pipelines. The input data are received through the AGILE data flow, described in Section \ref{sec:intro}, and saved in a Local Data Archive available for the analyses. In addition to the AGILE data, the notices received through the GCN network and managed by the GCN notices Receiver component are used as input. 

The AGILE pipelines are developed using the RTApipe framework \cite{2021ADASSParmiggiani}. This framework is designed to help developers of RTA pipelines by providing several features that satisfy the following requirements: (i) execute a great number of analyses exploiting parallel processing, (ii) allow the developers to configure different analysis workflows, (iii) perform different analyses for each type of science alerts or instruments that triggered it (e.g. GW, neutrino, Fermi/GBM etc.), and (iv) perform analyses with different science tools on several time scales and different configurations.

The RTA pipelines that can be developed with the RTApipe framework are divided into two categories: (i) the Archive Pipeline that analyses the collected data into the Local Data Archive as soon as they are received, and (ii) the Science Alert Pipeline that react to science alerts received by the GCN network and start a follow-up searching for a counterpart. The results of both pipelines are saved in different storage systems (MySQL and file system). In addition, a password-protected web Graphical User Interface (GUI), called Control Room, can be used by the AGILE Team to visualise the analysis results. 

The RTApipe framework uses the Slurm (\url{https://slurm.schedmd.com}) workload manager to execute analyses in parallel and manage priority between processes. The software and services are installed in a Singularity container (\url{https://sylabs.io}) that can be easily deployed on different hardware and operating systems.

\section{AGILE real-time analysis pipelines \label{sec:agilepipe}}

The AGILE Team developed three RTA pipelines, one of which is a Science Alert Pipeline. It reacts to internal or external science alerts (e.g. GRBs and GWs) and is described in Section \ref{subsec:agile_rta_pipe}. The other two pipelines are Archive Pipelines that perform a blind search of transient events (mainly GRBs) into the AGILE/MCAL data (Section \ref{subsec:mcal_rta}) and the ratemeters of all AGILE detectors (Section \ref{subsec:rm_rta}). The ratemeters contain the raw information about the acquisition event rate over time (e.g. the number of photons or particles detected per second) of all the detectors onboard AGILE and can be used to identify transient events. 

\subsection{Science Alert Pipeline \label{subsec:agile_rta_pipe}}

The Science Alert Pipeline is designed to respond to internal or external alerts and run an analysis of the data acquired from detectors onboard the AGILE satellite. The internal alerts are generated by the AGILE Archive Pipelines when a transient event is detected (Section \ref{subsec:mcal_rta}, \ref{subsec:rm_rta}), and they trigger the search for a transient counterpart in other AGILE detectors data (e.g. a science alert received by AGILE/MCAL triggers analysis on AGILE/GRID data). When a science alert is received, the pipeline sends a notification to the AGILE Team (SMS, email, and call). The AGILE Team can then follow up the science alerts visualising the pipeline's results.  

The Science Alert Pipeline performs two main types of analyses: (i) a prompt analysis performed with data of all the detectors onboard AGILE, using a time window centred in the transient event time to check if a counterpart detection is found near the transient event time and in the same sky position and (ii) a complete analysis (hereafter "full") to extract the scientific results in a larger time window ($\pm 1000$ seconds). The analyses performed for a single science alert are more than 100 and are executed with 20 different science tools that use the AGILE/GRID, AGILE/MCAL, and ratemeters data. Slurm manages the execution of these analyses in parallel to obtain the results as soon as possible, suspending low-priority jobs and starting the high-priority jobs related to the science alert. In a few seconds or minutes (depending on the task) since the arrival of the alert, the researchers on-duty can access the results using the web GUI.

\subsection{AGILE/MCAL pipeline \label{subsec:mcal_rta}}

The MCAL detector has a dedicated RTA pipeline that automatically searches for GRBs and Terrestrial Gamma-ray Flashes (TGF) \cite{2014JGRA..119.1337M} at each contact received by the satellite, storing the results in a MySQL database and in the file system. When a GRB is detected, the pipeline sends a notice to the GCN Network to notify the scientific community about this transient event. The algorithm detects GRBs independently from the MCAL trigger times, performing a blind search on different timescales (16 ms, 32 ms, 64 ms, and 128 ms), and identifying events that can be confidently neglected as statistical fluctuations of the background, where the background level is independently evaluated for each timescale. Whenever more than one bin exhibit a count rate N exceeding a given threshold level (which can be 3$\sigma$, 4$\sigma$, or 5$\sigma$, depending on the timescale under analysis), a candidate GRB is identified.

\begin{figure*}[!htb]
	\centering
	  \includegraphics[width=\linewidth]{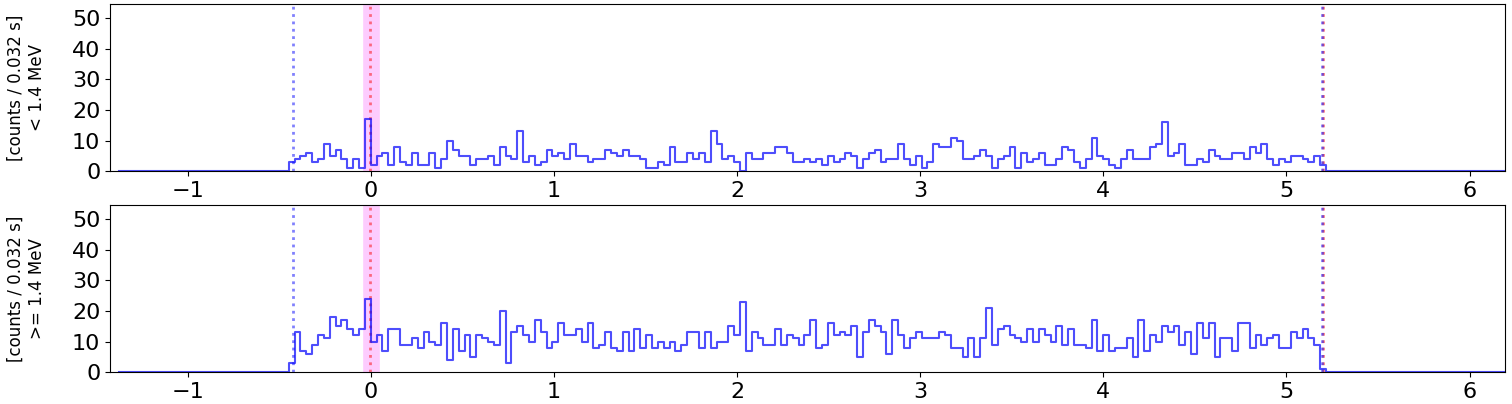}
	\caption{MCAL light curves with a possible burst highlighted with a magenta box.}
	\label{fig:mcal_gui}
\end{figure*}

The table of all the GRBs detected since the pipeline's first run is accessible within the Control Room and includes specific information for each GRB (e.g. the event signal-to-noise ratio, the counts, the background level, etc.). Moreover, it is possible to access each specific event on a web page showing the GRB's light curve (Fig. \ref{fig:mcal_gui}) representing the number of photon counts at different energy ranges.

The web page including the details for each TGF allows the user to visualise a two-panel plot reporting the event light curve and the energy of each photon detected, and a plot of the Earth equatorial surface (Earth map) with the satellite position marked by a red dot.

\subsection{AGILE ratemeters pipeline \label{subsec:rm_rta}}

\begin{figure*}[!htb]
	\centering
	  \includegraphics[width=\linewidth]{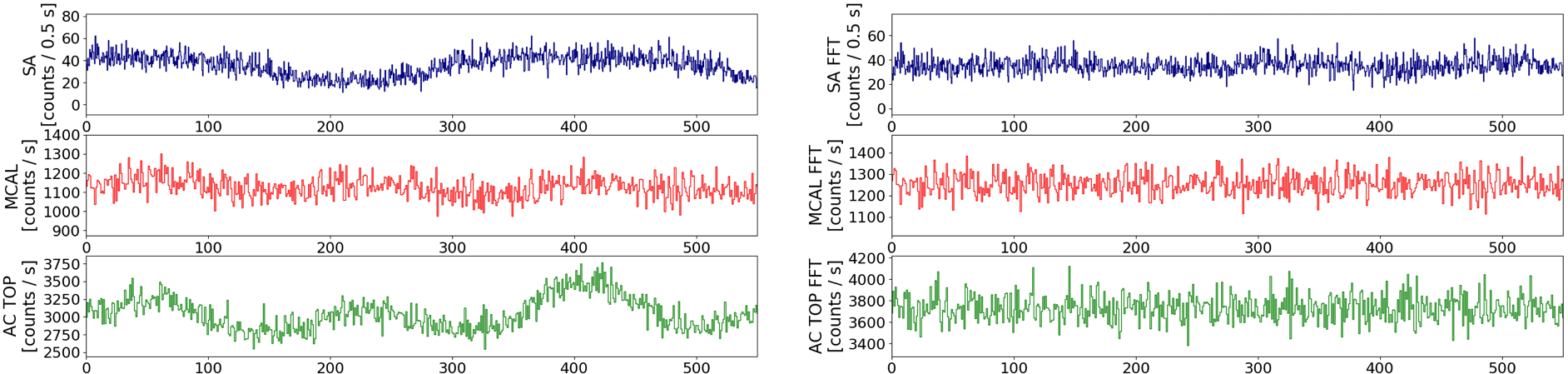}
	\caption{Light curves of the AGILE instrument ratemeters. The original data is shown in the left panels, while the right panels show the light curves after the trend removal.}
	\label{fig:rm_gui}
\end{figure*}

The AGILE Team developed an RTA pipeline to analyse the ratemeters (RM) acquired by all detectors onboard the AGILE satellite to detect fast transients, GRBs, and solar flares. The Control Room has a web page dedicated to the results of this pipeline for each contact. Fig. \ref{fig:rm_gui} shows multiple panel light curves covering part of a single contact data. Each panel represents the RM related to a detector onboard the satellite and shows the original light curves (left panels) and the detrended light curves after applying the Fourier transform to remove orbital and spinning modulations (right panels). The RM plots show the data of the following detectors: SuperAGILE, MCAL, and the AGILE Anticoincidence system described in Section \ref{sec:intro}.

The astronomer on-duty is in charge of monitoring the automated RM pipeline results during day and night, searching for transient events. According to pre-defined thresholds, an automated algorithm highlights with coloured boxes the candidate bursts inside the plots. 

\subsection{Results obtained with the AGILE RTA pipelines \label{subsec:rta_results}}

Since May 2019, the AGILE/MCAL pipeline was enabled to automatically send notices to the GCN network after a GRB detection. Until now, it sent more than 50 notices about GRBs containing the GRB's parameters and the light curves (\url{https://gcn.gsfc.nasa.gov/agile\_mcal.html}). 

In addition, the AGILE Team sent more than 150 circulars to the GCN network about the results obtained with the RTA pipelines during the follow-up of transient events detected by AGILE or by other facilities (in particular during the third LIGO/Virgo observing campaign) and several ATels communications. 

The AGILE Team published several papers where part of the results are obtained using the RTA pipelines, presented in this contribution, as a support system during the data analysis. The main papers are:  \cite{2021NatAs...5..401T}, \cite{2020ApJ...904..133U}, \cite{2020ApJ...890L..32C},  \cite{2019RLSFN.tmp...33V} and \cite{2019ApJ...870..136L}.

\section{Conclusions}\label{sec:conclusion}

This contribution presents the RTA pipelines developed for the AGILE space mission in MM and MW astronomy. These pipelines are implemented using the RTApipe framework that provides all the features to satisfy the software requirements. The main scenarios covered by the pipelines are two: (i) analyse the data acquired by the detectors onboard AGILE (GRID, MCAL) and in the scientific ratemeters as soon as they are available in the INAF/OAS Bologna data centre, and (ii) react to science alerts shared with the community by other observatories by performing a follow-up of transient events (e.g. GRBs, GWs, and neutrinos). The data analysis procedure is fully automated in both scenarios and does not require human intervention to generate results. In addition, the pipelines send email notifications to the AGILE Team when a transient event is detected and notices to the GCN network when the transient event is a GRB.

The software that performs automated data analysis is necessary to generate scientific results and detect transient events with low latency (seconds or minutes depending on the analysis) since the arrival of the alert. The AGILE Team organises personnel shifts to have at least a team member (astronomer on-duty) available to monitor the pipeline results during the day and the night using the web GUI to perform the follow-up of transient events detected by AGILE or by other observatories. Several papers, GCN circulars and notices have been published using the support of the RTA pipelines. The fast follow-up of transient events could not be possible without an automated analysis system.

\acknowledgments
The AGILE Mission is funded by the Italian Space Agency (ASI) with scientific and programmatic participation by the Italian National Institute for Astrophysics (INAF) and the Italian National Institute for Nuclear Physics (INFN). The investigation is supported by the ASI grant  I/028/12/6. We thank the ASI management for unfailing support during AGILE operations. We acknowledge the effort of ASI and industry personnel in operating the  ASI ground station in Malindi (Kenya), and the data processing done at the ASI/SSDC in Rome: the success of AGILE scientific operations depends on the effectiveness of the data flow from Kenya to SSDC and the data analysis and software management.



\bibliographystyle{JHEP}
\bibliography{proceeding}

%
%
%

\end{document}